\documentclass{article}
\usepackage[utf8]{inputenc}
\usepackage{url}
\usepackage{graphicx,graphics}
\usepackage{appendix}

\title{\vspace{-1cm}Compression is Comprehension: The Unreasonable Effectiveness of Digital Computation in the Natural World\thanks{Invited contribution to Chaitin's festschrift based on an invited talk delivered at the Workshop on \textit{Patterns in the World}, Department of Philosophy, University of Barcelona on December 14, 2018. Author's email: hector.zenil [at] algorithmicnaturelab [dot] org}}
\author{Hector Zenil$^{1,2,3}$\\
$^1$Oxford Immune Algorithmics,\\Oxford University Innovation, Oxford, U.K.;\\
$^2$Algorithmic Dynamics Lab\\Unit of Computational Medicine, Center for Molecular\\Medicine, Karolinska Institute, Stockholm, Sweden; \\
$^3$Algorithmic Nature Group, LABORES for the Natural\\and Digital Sciences, Paris, France.}

\date{}

\begin{document}

\maketitle

\begin{abstract}
Chaitin's work, in its depth and breadth, encompasses many areas of scientific and philosophical interest. It helped establish the accepted mathematical concept of randomness, which in turn is the basis of tools that I have developed to justify and quantify what I think is clear evidence of the algorithmic nature of the world. To illustrate the concept I will establish novel upper bounds of algorithmic randomness for elementary cellular automata. I will discuss how the practice of science consists in conceiving a model that starts from certain initial values, running a computable instantiation, and awaiting a result in order to determine where the system may be in a future state---in a shorter time than the time taken by the actual unfolding of the phenomenon in question. If a model does not comply with all or some of these requirements it is traditionally considered useless or even unscientific, so the more precise and faster the better. A model is thus better if it can explain more with less, which is at the core of Chaitin's ``compression is comprehension". I will pursue these questions related to the random versus possibly algorithmic nature of the world in two directions, drawing heavily on the work of Chaitin. I will also discuss how the algorithmic approach is related to the success of science at producing models of the world, allowing computer simulations to better understand it and make more accurate predictions and interventions.\\

\noindent \textsc{Keywords:} model-driven science; algorithmic randomness; Kolmogorov-Chaitin complexity; algorithmic information theory; mechanistic models; patterns.
\end{abstract}

\section{Introduction}

\begin{figure}[ht!]
\centering
\includegraphics[width=6cm]{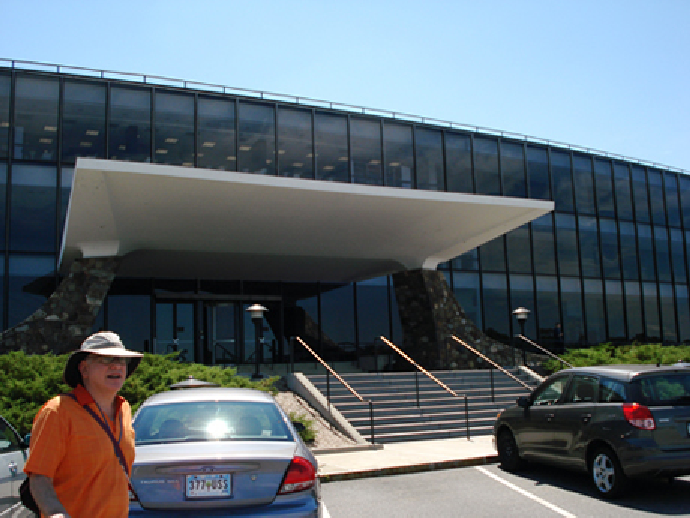}\\

\bigskip

\includegraphics[width=3.8cm]{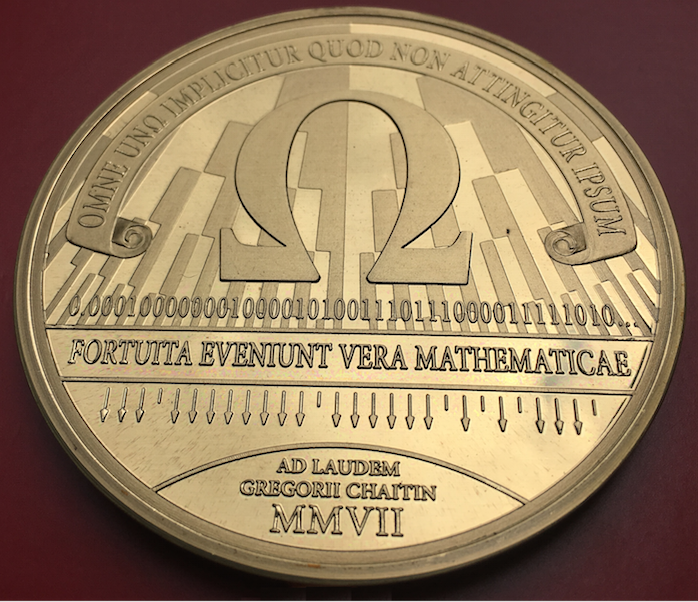}\hspace{.7cm}\includegraphics[width=4.05cm]{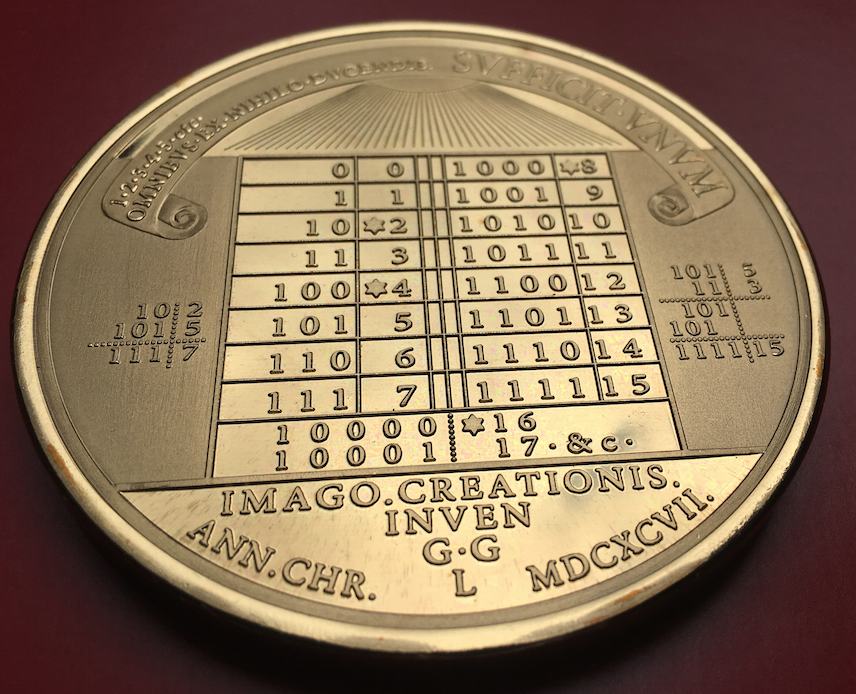}\\

\caption{(Top) A picture of Greg Chaitin I took outside his longtime office at IBM Research headquarters, the Thomas J. Watson Research Center. (Bottom) The two sides of the medal I helped Wolfram design, featuring material relating to Chaitin's life achievements on the obverse and Leibniz' own medal celebrating binary arithmetic on the reverse.}
\label{medal}
\end{figure}

The work of Greg Chaitin has been at the centre of my intellectual interests. He was not only one of the examiners for my PhD in computer science in 2011, but my work would have not been possible without his seminal ideas. I met Greg in 2006, having been invited to his home and to his office at IBM's legendary Thomas J. Watson Research Center (see Fig.~\ref{medal}), both in Yorktown Heights, NY, U.S. His house was typical of an intellectual, full of objects and books piled one on top of the other. But Greg also had exotic pieces of sculpture scattered around his house, and an over-representation of Leibniz- and G\"odel-related books, the two major sources of influence on his work and thinking.

\begin{figure}[ht!]
\centering
\includegraphics[width=8cm]{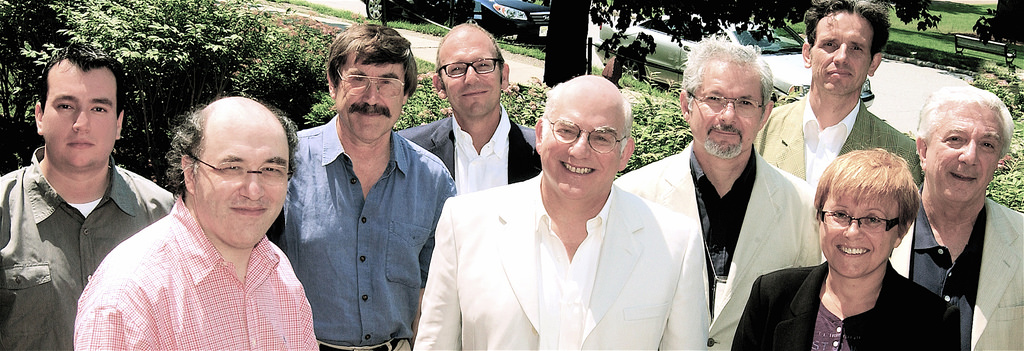}\\
Panel Part I: Is the universe random?\\

\bigskip

\includegraphics[width=8cm]{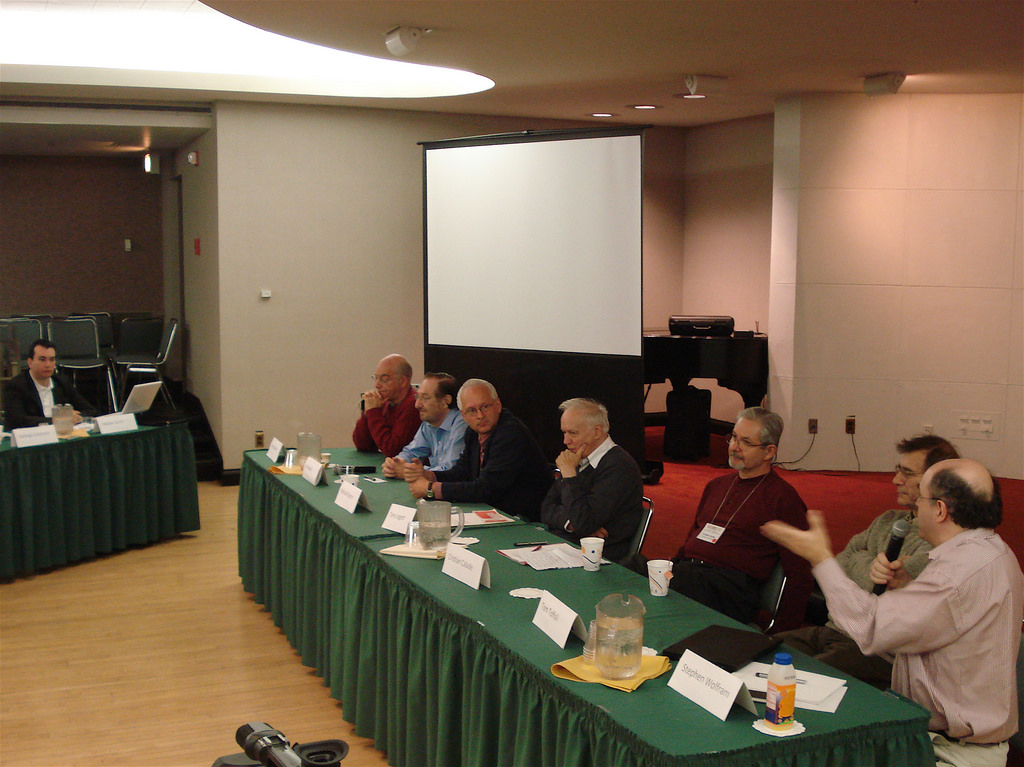}\\
Panel Part II: What is computation and How does nature compute?\\

\caption{Part I and Part II panel discussion pictures. (Top) From left to right: Hector Zenil, Stephen Wolfram, Paul Davies, Ugo Pagallo, Greg Chaitin, Cris Calude, Karl Svozil, Gordana Dodig Crnkovic, and John Casti. (Bottom) From left to right: Hector Zenil, Greg Chaitin, Ed Fredkin, Rob de Ruyter, Tonny Legett, Cris Calude, Tom Toffoli, and Stephen Wolfram. Transcriptions in ~\cite{randomness} and ~\cite{computableuniverse}.}
\label{panels}
\end{figure}

In 2008, I organized a two-part panel discussion (Fig.~\ref{panels}) during the Wolfram Science conference at the University of Vermont in Burlington, an event that, together with a second meeting that I organized (with Adrian German) in 2008 at Indiana University Bloomington, will, I believe, come to be regarded as events of significant historical value in the discussion of the ideas of the late 20th and early 21st centuries around the questions of the analogue and the digital. The first part of the panel discussion addressed the question ``Is the Universe random?" Participating were Cris Calude, John Casti, Greg Chaitin, Paul Davies, Karl Svozil and Stephen Wolfram. The second part addressed the question ``What is computation and How does nature compute?". The participants were Ed Fredkin, Greg Chaitin, Tonny Legett, Cris  Calude, Rob de Ruyter, Tom Toffoli, and Stephen Wolfram, with George Johnson, Gerardo Ortiz and myself serving as moderators. Transcriptions of both parts were published in my volume~\cite{randomness} and ~\cite{computableuniverse}.

In 2007, I also helped Stephen Wolfram design a commemorative medal to celebrate the 60th birthday of Gregory Chaitin, which also involved minting for the first time a 300 year old medal originally designed by Gottfried Leibniz to celebrate the invention or discovery of binary arithmetic (see Fig.~\ref{medal}). I published a blog post soon after we came up with the idea for the medal, explaining the pre- and 
post-minting story of the medal~\cite{blogpost}.

At the centre of my research is Greg Chaitin's work on what is known as algorithmic probability, which in turn is related to  Chaitin's Omega ($\Omega$) number, also called the halting probability. My very first research paper~\cite{zenilcalude} was published in Chaitin's 60th birthday festchrift~\cite{calude}. The paper advanced the claim and provided the first numerical evidence of a bias present in nature that I later advanced in a contribution that won an FQXi prize~\cite{zenilessay} on the digital or analogue nature of the universe. Such claims have more recently been noticed and expanded~\cite{kamal}, a further confirmation in which my own methods played an important role again. In a later follow-up piece, I connected Turing patterns with Turing machines, explaining how symmetry-breaking creates structure---as opposed to randomness---out of nothing, with only computability being assumed~\cite{naturalcomputing}.

\section{Complexity from Computation}

Not only did Chaitin help found one of the most exciting areas of modern science (computer science), but it may turn out that his contribution, together with that of Alan Turing, may have had a more profound effect on our understanding of our physical reality than we had hitherto supposed. 

At the beginning of the twentieth century and through the end of the Second World War, computers were human, not electronic, mainly women. The work of a computer consisted precisely in solving tedious arithmetical operations with paper and pencil. This was looked upon as work of an inferior order.
At an international mathematics conference in 1928, David Hilbert and Wilhelm Ackermann suggested the possibility that a mechanical process could be devised that was capable of proving all mathematical assertions. This notion is referred to as \textit{Entscheidungsproblem} (in German), or `the decision problem'. If a human computer did no more than execute a mechanical process, it was not difficult to imagine that arithmetic would be amenable to a similar sort of mechanization. The origin of Entscheidungsproblem dates back to Gottfried Leibniz, who having (around 1672) succeeded in building a machine based on the ideas of Blaise Pascal that was capable of performing arithmetical operations (named
Staffelwalze or the Step Reckoner), imagined a machine of the same kind that would be capable of manipulating symbols to determine the truth value of mathematical principles. To this end
Leibniz devoted himself to conceiving a formal universal language, which he designated characteristica universalis, a language that would encompass, among other things, binary language and the definition of binary arithmetic.

In 1931, Kurt G\"odel~\cite{godel} arrived at the conclusion that Hilbert's intention (also referred to as
`Hilbert's programme') of proving all theorems by mechanizing mathematics was not possible under certain reasonable assumptions. G\"odel advanced a formula that codified an arithmetical
truth in arithmetical terms and that could not be proved without arriving at a contradiction. Even worse, it implied that there was no set of axioms that contained arithmetic free of true formulae that could not be proved.

In 1944, Emil Post, another key figure in the development of the concepts of
computation and computability (focusing especially on the limits of computation) found~\cite{post} that this problem was intimately related to one of the twenty-three problems (the tenth) that Hilbert, speaking at the Sorbonne in Paris, had declared the most important challenges for twentieth century mathematics.

Usually, Hilbert's programme is considered a failure, though in fact it is anything but. Even though it is true that G\"odel debunked~\cite{godel} the notion that what was true could be proved, presenting a negative solution to the `problem of decision', and Martin Davis~\cite{davis} (and independently, Julia Robinson~\cite{robinson}) used G\"odel's negative result to provide a negative solution to Hilbert's tenth problem (the argument for which was completed by Yuri Matiyasevich~\cite{matiyasevich}), Hilbert's supposedly failed programme originated what we now know as Computer Science, the field that wouldn't have been possible without Alan M. Turing's concept of the universal machine.

\subsection{One machine for everything}

Not long after G\"odel, Alan M. Turing made his appearance. Turing contemplated the problem of decision in much cruder terms. If the act of performing arithmetical operations is mechanical, why not substitute a mechanical device for the human computer? Turing's work represented the first abstract description of the digital general-purpose computer as we know it
today. Turing defined what in his article he termed an `a' computer (for `automatic'), now known as a Turing machine.

A Turing machine is an abstract device which reads or writes symbols on a tape one at a time, and can change its operation according to what it reads, and move forwards or backwards
through the tape. The machine stops when it reaches a certain configuration (a combination of what it reads and its internal state). It is said that a Turing machine produces an output if the
Turing machine halts, while the locations on the tape the machine has visited represent the
output produced.

The most remarkable idea advanced by Turing is his demonstration that there is an `a' machine that is able to read other `a' machines and behave as they would for an input s. In other words, Turing proved that it was not necessary to build a new machine for each different task; a single machine that could be reprogrammed sufficed. This erases the distinction between program and data, as well as between software and hardware, as one can always codify data as a program to be executed by another Turing machine and vice versa, just as one can always build a \textit{universal machine} to execute any program and vice versa.

Turing also proved that there are Turing machines that never halt, and if a Turing machine is to be \textit{universal}, and hence able to simulate any other Turing machine or computer program, it is actually expected that it will never halt for an infinite number of inputs of a certain type (while halting for an infinite number of inputs of some other type). And this is what
Turing would have expected, given G\"odel's results and what he wanted to demonstrate: that Hilbert's mechanization of mathematics was impossible. This result is known as the undecidability of the halting problem.
In his seminal article Turing defined not only the basis of what we today know as digital general-purpose computers, but also software, programming and subroutines. And thus without a doubt it represents the best answer to date that we have to the question `What is an algorithm?'

In fact in Alan Turing's work on his universal machine, he even introduced the concept of a subroutine that helped him in his machine construction. These notions are today the cornerstone of the field that Turing, more than anyone else, helped establish, viz. Computer Science.

Once we approach the problem of defining what an algorithm is and arrive at the
concept of \textit{universality} that Turing advanced, the question to be considered in greater detail concerns the nature of algorithms. Given that one now has a working definition of the algorithm, one can begin to think about classifying problems, algorithms and computer programs by, for example, the time they take or the storage memory they may require to be executed. One may assume that the time required for an algorithm to run would depend on the type of machine, given that running a computer program on a Pentium PC is very different from executing it on a state-of-the-art super computer. This is why the concept of the Turing machine
was so important---because any answers to questions about problem and algorithm resources will only make sense if the computing device is always the same. And that device is none other than the universal Turing machine. So for example, every step that a Turing machine performs while reading its tape is counted as a time step.
Many algorithms can arrive at the same conclusion taking different paths, but some may be faster than others, but this is now a carefully considered matter when fixing the framework
for Turing's model of computation: one asks whether there is an algorithm that surpasses all others in economy as regards the resources required when using exactly the same computing
device. These are the questions that opened up an entire new field in the wake of Turing's work,
the development of which Turing would certainly have been delighted to witness. This field is today referred to as the theory of Computational Complexity, which would not have been possible without a concept such as that of the universal Turing machine. The theory of Computational Complexity focuses on classifying problems and algorithms according to the
time they take to compute when larger inputs are considered, and on how size of input and execution time are related to each other. This is all connected to two basic resources needed in
any computation: space (storage) and time. For example, one obvious observation relating to this theory is that no algorithm will need more space than time to perform a computation. One
can then quickly proceed to ask more difficult but more interesting questions, such as whether a machine can execute a program faster if it is allowed to behave probabilistically instead of  deterministically. What happens if one adds more than one tape to a universal Turing machine operation? Would that amount to implementing an algorithm to solve a problem much faster? Or one may even ask whether there is always an efficient algorithm for every inefficient algorithm, a question that may lead us to a fascinating topic connecting computers and physics.

\subsection{The world of simple programs}

If algorithms can come in a variety of types, some slow, others faster, what is it that allows nature to produce, with no apparent effort, what seem to us to be complex objects? These range from the laws of physics to the formation of matter and galaxies to the beginning of life on Earth (and possibly in other parts of the universe). In the end, one can see all these natural phenomena as a kind of computation, regardless of whether it is of exactly the same type as that performed by a Turing machine. This latter possibility cannot be completely disregarded. Thanks to Turing we know that even simple devices such as universal Turing machines possess incredible power.

One of the natural world's most fascinating characteristics is that it presents a wide range of physical and biological systems that behave in different ways, just like algorithms,
most of them having some regular features while nonetheless being hard to predict. Climate is a case in point. Even though it is cyclical, it is impossible to predict its details more than a week in advance. Where does nature's complexity come from? Throughout human history we have encountered objects, in particular mathematical ones, that seem complex to us. One set of such objects comprises numbers that can be expressed as the division $p/q$, with $p$ and q being integers. Numbers 5, 0.5 or even infinite numbers such as 0.333… can be written as 5/1, 1/2, and 1/3, respectively. But as far back as the ancient Greeks numbers have been known, such as $\pi$ and the square root of 2, which cannot be expressed in this way. One could think of arithmetical division as an algorithm that can be performed by a Turing machine, the result being provided in the output tape. Multiplication, for example, is an algorithm to shorten the number of steps needed to perform the same operation using only addition. In the case of numbers that admit a rational representation $p/q$, the algorithm of the division of integers
consists of the common procedure of finding quotients and remainders. In the case of numbers such as $\pi$ and the square root of 2, the algorithm produces an infinite non-periodic expansion, so
that the only way to represent them is symbolically (i.e. $\pi$ and $\sqrt{2}$). The Pythagoreans found that
those numbers with ostensible infinite complexity could be produced from very simple operations, for example, when seeking the value of the hypotenuse of a right triangle with sides of length 1. Since Euclid, it has also been known that such numbers are not the exception among real numbers that are found, for example, in the continuous interval (0, 1).

In algorithmic terms, rational and irrational numbers are different in nature. When one starts a Turing machine that implements the algorithm for division, there is no algorithm that
allows for the production of an irrational number followed by halting, whereas the division of rational numbers can halt (when the remainder is zero) or enter an infinite cycle that will
produce a repetitive decimal expansion.

In engineering, including systems programming, the intuition of what is complex (in comparison to an irrational number in mathematics, for example) has been radically different.

The usual assumption has been that to produce something that seems complex, a process that is just as complex has to be devised. This issue, however, is closely connected to Turing's concept
of universality, given that a universal Turing machine that is programmable is, in principle, capable of producing any degree of `complexity', for example, the type of complexity (or randomness) that one can see in the decimal expansion of $\pi$.

If Euclid's algorithm for division or $\pi$ can produce such apparent complexity, how usual is it to run a random computer program that produces the same complexity? If computer
programs that produce complexity need a very complex description, the probability of finding one small enough would be very low. For example, even though Turing's 1936 article contains
all the main elements of the traditional description of a universal Turing machine capable of reproducing the type of complexity to be found in the digits of $\pi$, the construction of his
universal machine requires at least 18 states, and at least 23 instructions (the exact number cannot be calculated on the basis of Turing's article due to the fact that he uses subroutines that
can be implemented on machines of different sizes).

Whatever the actual threshold for reaching Turing universality, it had typically been thought to be high (a case in point: von Neumann's universal builder, a system that was the
anticipation of the modern concept of cellular automata, requires 29 states), and it was thought that a universal machine would require a certain minimum complexity (at least as to the number
of states and symbols required to describe it). In an experiment with extremely small and simple
computer programs, Stephen Wolfram found that this threshold of complexity and universality was likely to be extremely low~\cite{wolfram}, and that very little was required to find a machine that produced high complexity or that was capable of being Turing universal. Indeed, not only did Wolfram find a very small Turing machine with only 2 symbols and 5 states that was capable of carrying out universal computation under some very simple conditions, that is, a computer that was powerful enough to emulate any standard computer, 
there was another Turing machine that Wolfram suspected was Turing-universal, and to ascertain whether this was so, in 2007 Todd Rowland and I organized a prize competition  (\url{https://www.wolframscience.com/prizes/tm23/}) with a view to determining how simple the rules for a universal Turing machine could in fact be.

Wolfram's other favourite computer programs are called Elementary Cellular Automata~\cite{wolfram} and it is on one of these that the 2,5 original smallest universal Turing machines are based.

Elementary Cellular Automata (ECA) are another superb example of a computing model capable of very rich behaviour. ECA are minimalistic computer programs that are very visual, as they print out their space-time evolution in two dimensions and can be subjected to easy visual inspection. Recently, my colleagues and I found another extremely small Turing-universal computer by combining two ECA, which are themselves extremely simple~\cite{zeniljurgen}.

In Fig.~\ref{simplified}, I show how the ideas of Chaitin (and Kolmogorov and Solomonoff) can help understand the complexity of ECA by looking at how difficult it is to describe them succinctly from their generating model. The simplified rule shown in Fig.~\ref{simplified} is an upper bound on their Kolmogorov-Chaitin complexity. Such simple estimations already provide a much better characterization than other simplifications, such as that of the so-called Langton's $\lambda$ parameter, as they correspond better to the literature on the complexity of ECA. 
And even better estimations and tighter bounds can be found using more powerful approaches based on Chaitin's (and Levin's) work, notably with the help of two methods that I and my team put together called CTM and BDM 
(as shown in Fig.~\ref{simplified}), rooted in a beautiful concept called algorithmic probability, which is deeply related to Chaitin's own $\Omega$ number.

\subsection{Algorithmic probabilities}
\label{ctm}

Just as the formulae for the production of the digits of $\pi$ are compressed versions of $\pi$, the laws of physics can be seen as systems that compress natural phenomena. These laws are
valuable because it is thanks to them that the result of a natural phenomenon can be predicted without having to wait for it to unfold in real time, e.g. one can solve the equations that describe
planetary movement instead of waiting two years to know the future positions of a planet. For all practical purposes the laws of physics are like computer programs and the scientific models
we have of them: executable on digital computers and susceptible of numerical solutions.

At the centre of Chaitin's work related to his Chaitin $\Omega$~\cite{chaitin}, and also the variation introduced by Solomonoff~\cite{solomonoff} and Levin~\cite{levin}, is the notion of algorithmic probability, which can be defined as,

$$AP(s) = \sum_{p:U(p) = s} 1/2^{|p|} $$

That is, the sum over all the programs $p$ for which a universal Turing machine $U$ outputs $s$ and halts.

The notion behind $AP$ is very intuitive. If one wished to produce the digits of $\pi$ randomly, one would have to try time after time until one managed to hit upon the first numbers corresponding to an initial segment of the decimal expansion of $\pi$. The probability of success is extremely small: $1/10$ digits multiplied by the desired quantity of digits. For example, $1/10^{2400}$ for a segment of 2400 digits of $\pi$. But if instead of shooting out random numbers one were to
shoot out computer programs to be run on a digital computer, the result would be very different.

A program that produces the digits of $\pi$ would have a higher probability of being produced by a computer. Concise and known formulas for $\pi$ could be implemented as short computer programs that would generate any arbitrary number of digits of $\pi$.

\subsection{Algorithmic Chaitin complexity}

The length of the shortest program that produces a string is today the mathematical definition of randomness, as introduced by Kolmogorov~\cite{kolmo}, Chaitin~\cite{chaitin}, and Solomonoff~\cite{solomonoff}, and later expanded by Levin~\cite{levin}, and also called Kolmogorov-Chaitin complexity. 

The idea is relatively simple. If a string $s$ of length $|s|$ cannot be produced by a program $|p|$ shorter than $|s|$ in bits, then the string $s$ is random because it cannot be effectively described in a shorter way than by $s$ itself, there being no program $p$ that generates $s$ whose length is shorter than $s$. Formally,

$$C_U(s) = \min\{|p|:U(p)=s\}$$

\noindent where $U$ is a universal Turing machine. The invariance theorem~\cite{kolmo,chaitin,solomonoff} guarantees that for some universal Turing machines, the value of $C$, whether calculated with one particular universal Turing machine or another, $C$ can be bounded by a (small) constant. Formally, if $U_1$ and $U_2$ are two universal Turing machines and $C_{U_1}(s)$ and $C_{U_2}(s)$ are the values for the algorithmic complexity of $s$ for $U_1$ and $U_2$ respectively, there exists a constant $c$ such that

$$|C_{U_1}(s) - C_{U_2}(s)| < c$$

Thus, the longer the string, the less important $c$ is and the more stable the algorithmic complexity value $C$ is.
One of the disadvantages of $C$ is that, given the halting problem for Turing machines, $C$ is not computable, which is to say that given a string, there is no algorithm that returns the
length of the shortest computer program that produces it.

\subsection{Complexity is in inverse relation to probability}

Algorithmic probability and algorithmic complexity $K$ are inversely proportional, as established by the so-called algorithmic Coding theorem~\cite{cover,calude}:

$$|-\log_2 AP(s) - C(s)| < c $$

\noindent where $c$ is a constant independent of $s$. The Coding theorem implies that the algorithmic complexity can be estimated from the frequency of a string.

To illustrate the above let us consider $\pi$. Under the assumption of Borel's absolute normality of $\pi$, whose digits appear randomly distributed, and with no knowledge of the deterministic source and nature of $\pi$ as produced by short mathematical formulae, we ask how an entropy versus an algorithmic metric performs. First, the Shannon entropy rate (thus assuming the uniform distribution along all integer sequences of $N$ digits) of the $N$ first digits of $\pi$, in any base, would suggest maximum randomness at the limit. However, without access to or without making assumptions as regards the probability distribution, approximations to algorithmic probability would assign $\pi$ high probability, and thus the lowest complexity by the Coding theorem, as has been done in~\cite{d4,d5,kolmo2d,bdmpaper}.

Just as with $\pi$ but in application to graphs and networks, it has been proven how certain graphs can be artificially constructed to target any level of Shannon entropy~\cite{zkpaper,morzy}, preserving low algorithmic randomness.

But how relevant is the algorithmic Coding theorem in explaining, e.g., natural phenomena, if it only applies to Turing-universal systems? We know that the natural world allows and can carry out Turing-universal computation because we have been able to build computers that take elements from nature and make them perform as general-purpose computers. However, we don't know how much of the natural world is Turing-computable or how physical laws may naturally implement any form of computation. So, in~\cite{liliana} we showed that up to 60\% of the bias found in the output of systems that are not Turing universal may be explained by the algorithmic Coding theorem. This means that this theorem is far more relevant than expected, as it not only explains the way data and patterns distribute for a very particular computational model, but does so, to a considerable extent, for basically any computing model. It is not difficult to imagine that nature operates 
at some of these levels, perhaps not even on a fixed one but at multiple scales, with the algorithmic Coding theorem being relevant to all of them.

\subsection{The Coding Theorem Method}

The algorithmic \textit{Coding Theorem Method} (CTM)~\cite{d4,d5} provides the means for approximation via the frequency of a string. Now, why is this so? The underlying mathematics originates from the relation specified by algorithmic probability between frequency of production of a string from a random program and its algorithmic complexity. It is also therefore denoted as the algorithmic \textit{Coding theorem}, in contrast to another well known coding theorem in classical information theory. Essentially, the numerical approximation hinges on the fact that the more frequently a string (or object) occurs, the lower its algorithmic complexity. Conversely, strings with a lower frequency have higher algorithmic complexity. Otherwise stated,

$$CTM(s) = - \log_2 AP(s) + c$$

The way to implement a compression algorithm at the level of Turing machines, unlike popular compression algorithms, which are heavily based on Shannon entropy, is to go through all possible compression schemes. This is equivalent to traversing all possible programs that are a compressed version of a piece of data, which is exactly what the CTM algorithm does.

\subsection{The Block Decomposition Method}

Our approach to Chaitin's halting probability and Solomonoff-Levin's algorithmic probability consists in asking after the probability of a matrix being generated by a random Turing machine on a 2-dimensional array, also called a \textit{termite} or \textit{Langton's ant}~\cite{langton}. Hence an accounting procedure is performed using Turing machines that aims to approximate the algorithmic complexity of the identified structures. This technique is referred to as the \textit{Block Decomposition Method} (BDM), as introduced in~\cite{zenilgraph} and ~\cite{bdmpaper}. The BDM technique requires a partition of the adjacency matrix corresponding to the graph into smaller matrices. With these building blocks at hand we numerically calculate the corresponding algorithmic probability by running a large set of small 2-dimensional deterministic Turing machines, and then---by applying the algorithmic Coding theorem, as discussed above---its algorithmic complexity. 

Following such a divide-and-conquer scheme we can then approximate the overall complexity of the original adjacency matrix by the sum of the complexity of its parts. Note that we have to take into account a logarithmic penalization for repetition, given that $n$ repetitions of the same object only add $\log n$ to its overall complexity, as one can simply describe a repetition in terms of the multiplicity of the first occurrence. Technically, this translates into the algorithmic complexity of a labelled graph $G$ by means of $BDM$ which is defined as follows:

\begin{equation}
\label{newecaeq}
BDM(G,d) = \sum_{(r_u,n_u)\in A(G)_{d\times d}} \log_2(n_u)+K_m(r_u)
\end{equation}
where $K_m(r_u)$ is the approximation of the algorithmic complexity of the sub-arrays $r_u$ arrived at by using the algorithmic Coding theorem, while $A(G)_{d\times d}$ represents the set with elements $(r_u,n_u)$, obtained by decomposing the adjacency matrix of $G$ into non-overlapping squares, i.e. the block matrix, of size $d$ by $d$. In each $(r_u,n_u)$ pair, $r_u$ is one such square and $n_u$ its multiplicity (number of occurrences). From now on $BDM(G,d=4)$ will be denoted only by $BDM(G)$, but it should be taken as an approximation to $C(G)$ unless otherwise stated (e.g. when taking the theoretical true $C(G)$ value). Once CTM is calculated, BDM can be implemented as a look-up table, and hence runs efficiently in linear time for non-overlapping fixed size submatrices.

\subsection{Algorithmic Information Dynamics}

Unlike most complexity measures that are designed for static objects, except those related to dynamical systems (e.g. Lyapunov exponents), I have led the introduction of a measure of algorithmic complexity adapted for dynamical systems and designed to characterize the change of algorithmic complexity of an object evolving over time. The measure is universal in the sense that it can deal with any computable feature that the system may display over time, either spontaneously or as a result of an external perturbation/intervention/interaction.

At the core of Algorithmic Information Dynamics~\cite{maininfo,crc}, the algorithmic causal calculus that we introduced, is the quantification of the change of complexity of a system under natural or induced perturbations, particularly the direction (sign) and magnitude of the difference of algorithmic information approximations denoted by $C$ between an object $G$, such as a cellular automaton or a graph and its mutated version $G^\prime$, e.g. the flip of a cell bit (or a set of bits) or the removal of an edge $e$ from $G$ (denoted by $G\backslash e= G^\prime$). The difference $| C(G) - C(G\backslash e) |$ is an estimation of the shared algorithmic mutual information of $G$ and $G\backslash e$. If $e$ does not contribute to the description of $G$, then $| C(G) - C(G\backslash e) | \leq \log_2|G|$, where $|G|$ is the uncompressed size of $G$, i.e. the difference will be very small and at most a function of the graph size, and thus $C(G)$ and $C(G\backslash e)$ have almost the same complexity. If, however, $| C(G) - C(G\backslash e) | \leq \log_2|G|$ bits, then $G$ and $G\backslash e$ share at least $n$ bits of algorithmic information in element $e$, and the removal of $e$ results in a loss of information. In contrast, if $C(G) - C(G\backslash e) > n$, then $e$ cannot be explained by $G$ alone, nor is it algorithmically not contained/derived from $G$, and it is therefore a fundamental part of the description of $G$, with $e$ as a generative causal mechanism in $G$, or else it is not part of $G$ but has to be explained independently, e.g. as noise. Whether it is noise or part of the generating mechanism of G depends on the relative magnitude of n with respect to $C(G)$ and to the original causal content of $G$ itself. If $G$ is random, then the effect of $e$ will be small in either case, but if $G$ is richly causal and has a very small generating program, then $e$ as noise will have a greater impact on $G$ than would removing $e$ from an already short description of $G$. However, if $| C(G) - C(G\backslash e) | \leq \log_2 |G|$, where $|G|$ is, e.g., the vertex count of a graph, or the runtime of a cellular automaton, $G$, then $e$ is contained in the algorithmic description of $G$ and can be recovered from $G$ itself (e.g. by running the program from a previous step until it produces $G$ with $e$ from $G\backslash e$).

We have shown how we can infer and reconstruct space-time evolutions by quantification of the disruptiveness of a perturbation. We can then extract the generating mechanism from the ordered time indices, from least to most disruptive, and produce candidate generating models. Simpler rules have simpler hypotheses, with an almost perfect correspondence in row order. Some systems may look more disordered than others, but locally the relationship between single rows is for the most part preserved (indicating local reversibility). 

We have shown that the later in time a perturbation is injected into a dynamical system the less it contributes to the algorithmic information content of the overall space-time evolution. We then move from discrete 2D systems to the reconstruction of phase spaces and space-time evolutions of $N$-dimensional, continuous, chaotic, incomplete and even noisy dynamical systems.

\section{Computability in Nature}

Chaitin's work has allowed me to frame and attempt to answer questions related to the random or algorithmic nature of the world and universe: whether it is in some sense mechanistic and can be reproduced by an artificial mechanism such as an electronic computer. A first point to note is that we as human observers access the world through what we call science, at least when we do so formally, and the aim of science is to perform causal inference from empirical observations about the world or the universe. And the driver is twofold: to understand the world and to formulate predictions regarding natural phenomena. The result of this approach over the last millennia has been quite successful; some may describe it as incredible, even unreasonable.

\subsection{Cellular automata as inexhaustible pattern generators}

The cellular automata (CA) model can be seen as a pattern-generating computational system, and because the model is Turing-complete it is in a formal sense an inexhaustible source of complexity generation. CA have an intrinsic parallel nature that make them ideal for visually representing a computation. We recognize certain CA performing computation as having specific meanings, for instance, colliders, counters, and majority deciders.

\begin{figure}[ht!]
\centering

\textnormal{   }{\small Rule 255}\\

\medskip

\scalebox{.21}{\includegraphics{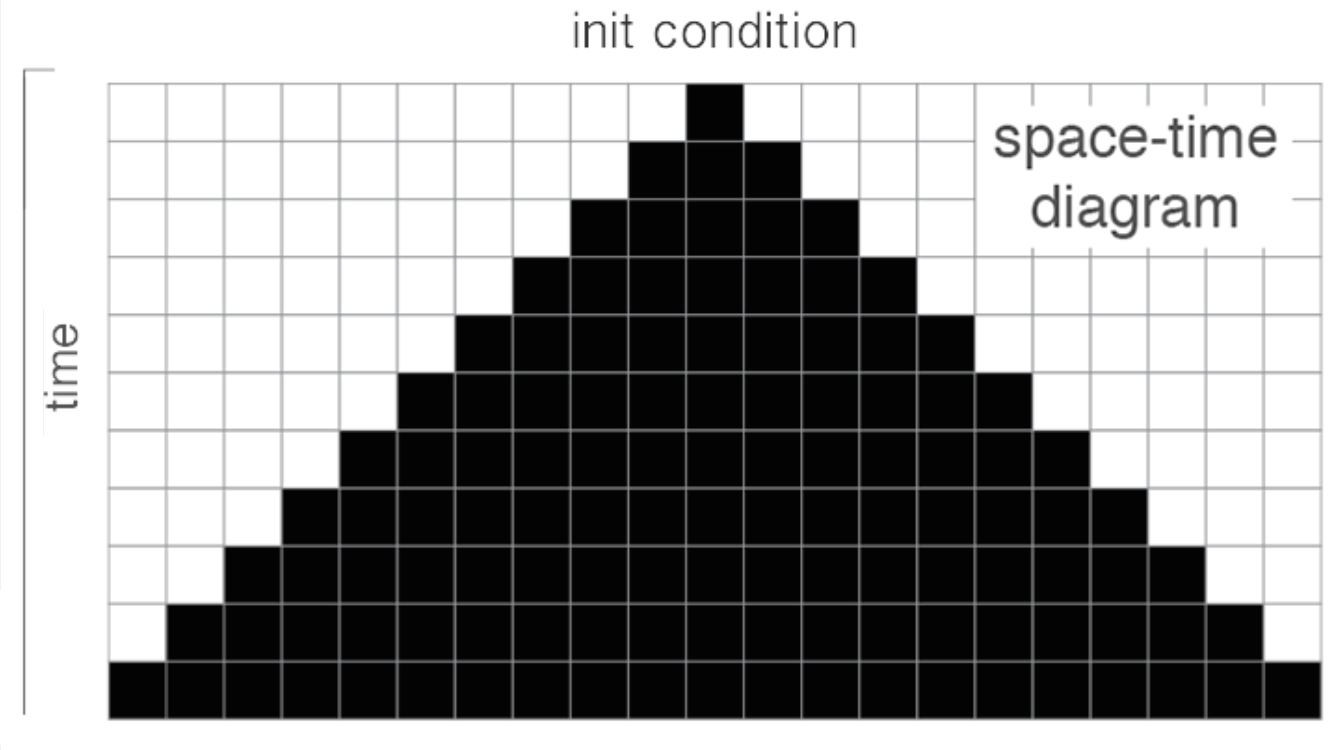}}

\bigskip
\bigskip

Estimations of Algorithmic Complexity for different ECA rules\\

\medskip

\scalebox{.3}{\includegraphics{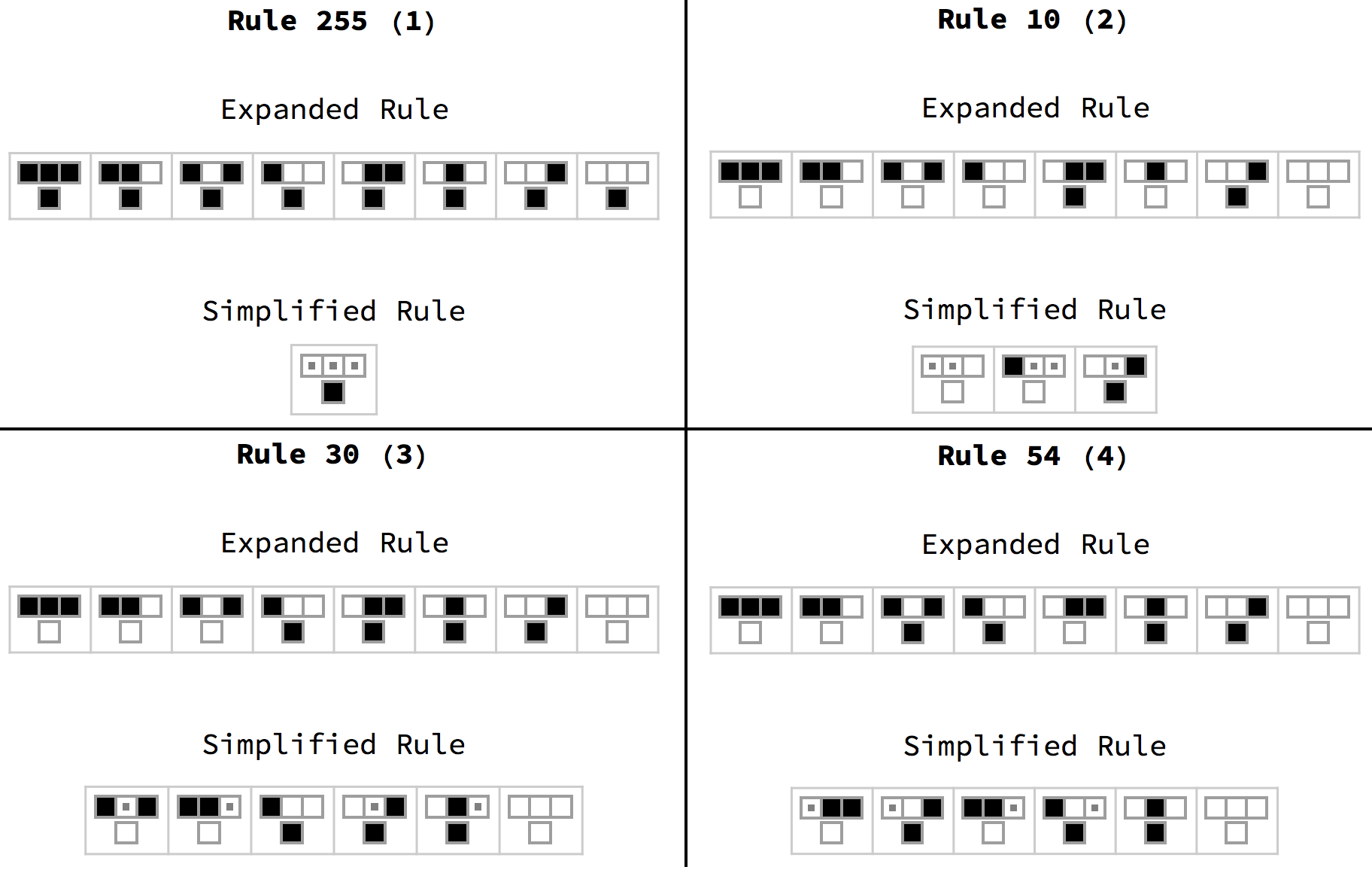}}

\bigskip
\bigskip

\caption{\label{simplified}(Top) An example of a cellular automaton, in this case Elementary Cellular Automaton (ECA) rule 255 which consists in traversing the cells three by three along each row and converting any three cells starting from the original initial condition (a single black cell) and printing another black cell below. (Middle) ECA cases showing the process of wildcard rule simplification. For rule 255 the simplified description of the ECA tells us that any black or white combination of three cells always leads to the same black cell.}
\end{figure}

To illustrate the concept of algorithmic complexity, we can represent ECA rules in a simplified fashion as shown in Fig.~\ref{simplified}, featuring the original rule and the simplified one using wildcards .

The use of `wildcards' allows some sort of trivial compression allowing a shorter description, and thus tighter upper bounds to Kolmogrov-Chaitin complexity as compared to their original, longer 8-bit descriptions. In Fig.~\ref{simplified} four ECA are shown, with their rules represented as sets of local rules or icons followed by their wildcard simplified version, which consists in reducing the description length of the global rule by looking at ways to compress the local rules. An implementation of this type of wildcard simplification can be found in the Wolfram Language under the RulePlot[] function with Appearance $\rightarrow$ ``Simplified" indicated. Clearly, the more random-looking and complex, the larger the wildcard description. Each rule allows a greater simplification (compared to its original 8-bit description) when the rule produces less algorithmically complex patterns (the number followed by each rule is its Wolfram class). 

\begin{figure}[ht!]
\centering
\scalebox{.38}{\includegraphics{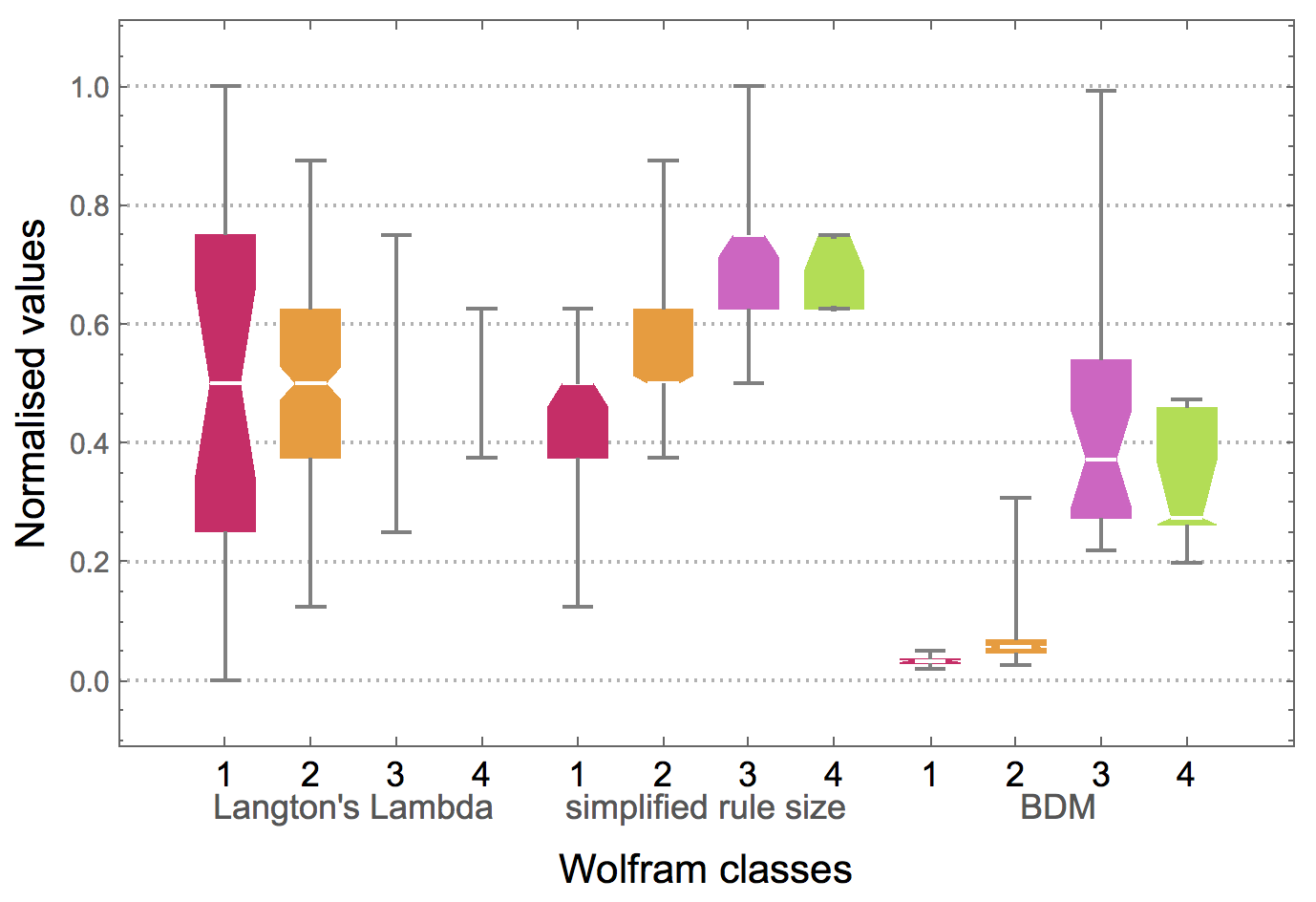}}
\caption{\label{ecas}An a priori measure based on non-zero density (called Langton's lambda or $\lambda$) of cellular automata versus tighter bounds of algorithmic randomness according to the rule simplification scheme, and one of the most popular lossless compression algorithms (LZW) used to approximate algorithmic complexity. Values are normalized by the maximum value of each index applied to all ECA starting from a random initial configuration and letting each rule evolve for 100 steps. LZW collapses all cases because they are too small to differentiate them, but even for larger runtimes LZW confounds most classes, particularly 1 with 2, and 3 with 4. In contrast, BDM assigns higher algorithmic randomness to class 3 compared than to class 4. BDM also collapses the simplest classes 1 and 2 more than simpler indexes, including the popular LZW used and abused to approximate algorithmic complexity.}
\end{figure}

The simplified rule size as described in Fig.~\ref{simplified} is an a priori algorithmic randomness measure based on a simplification of the CA global rule, and BDM is an a posteriori observer-oriented measure based Chaitin's randomness. Introduced by Wolfram~\cite{wolfram}, Wolfram's classes correspond to the typical behaviour of each ECA based on how often a rule behaves in a certain way for a random initial condition. Classes 1 and 2 are the most simple, while class 3 displays statistically-random behaviour and class 4 displays random behaviour together with sophisticated structures with which computation can be implemented~\cite{wolfram,cook}.
 
In Fig.~\ref{ecas}, it is clear how an a priori parameter such as Langton's $\lambda$ that measures the density of non-zeros in the description of a rule as a measure of complexity deals poorly with the system and is easily outperformed by methods based on Chaitin's algorithmic randomness, one of which (BDM) is numerically based on a variation of Chaitin's $\Omega$, unlike methods such as LZW and other popular lossless compression algorithms that are mostly based on Shannon entropy~\cite{liliana}. BDM quantifies algorithmic complexity circumventing the use of any popular lossless compression algorithm which are misleading as they are closer to Entropy than to algorithmic complexity~\cite{liliana}.

\subsection{Science: mapping data to models}

So what compression can tell us about the type of knowledge and the way in which we can access the world?

While the practice of science, and mechanistic models in particular, points towards an algorithmic universe governed by ordered laws and computable principles, it is only with the theory of algorithmic randomness that we are able to mathematically understand and quantify the subtleties in this space of computable models.

\begin{figure}[ht!]
\centering
\scalebox{.32}{\includegraphics{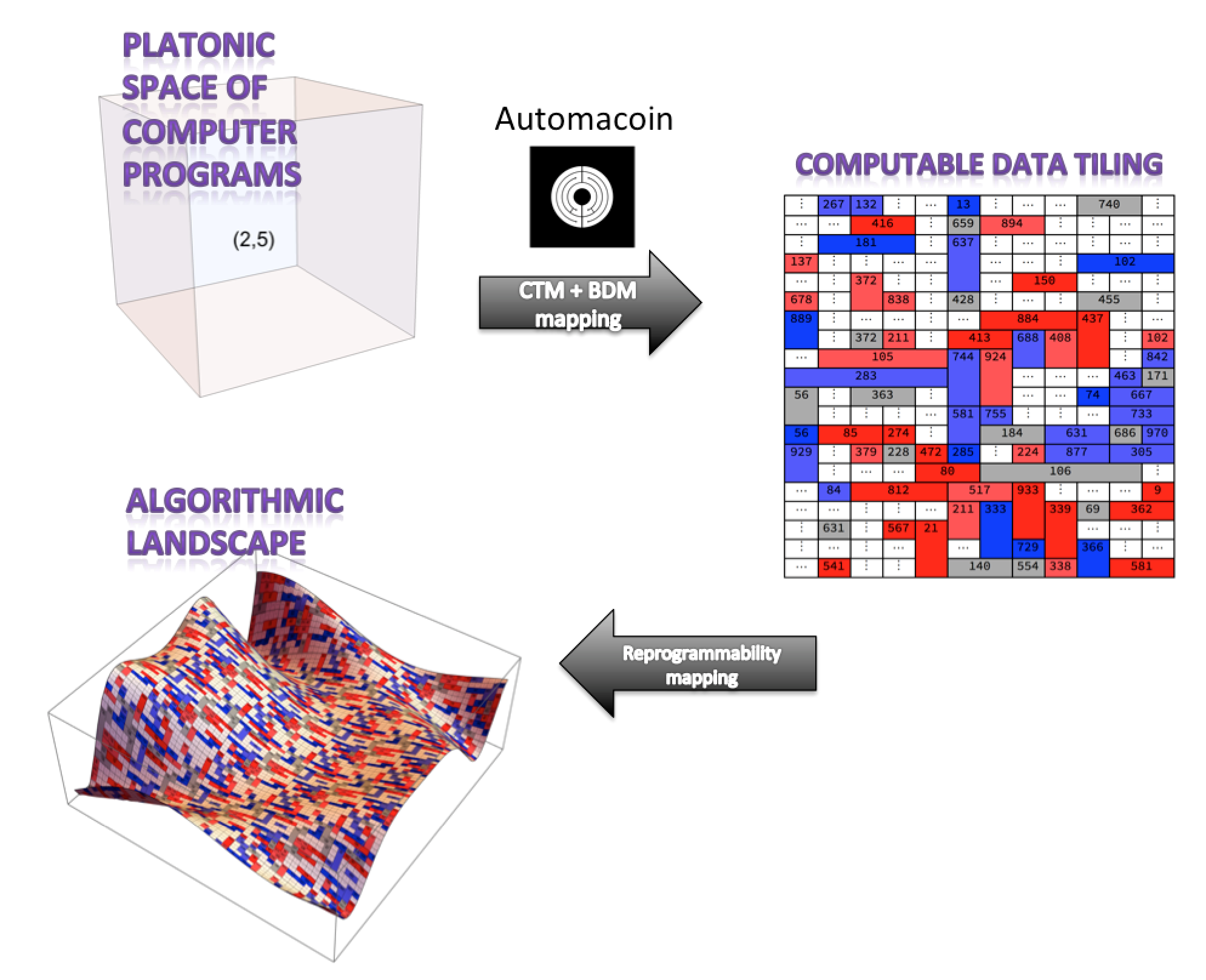}}

\bigskip

Coding Theorem and Block Decomposition Methods

\medskip

\scalebox{.39}{\includegraphics{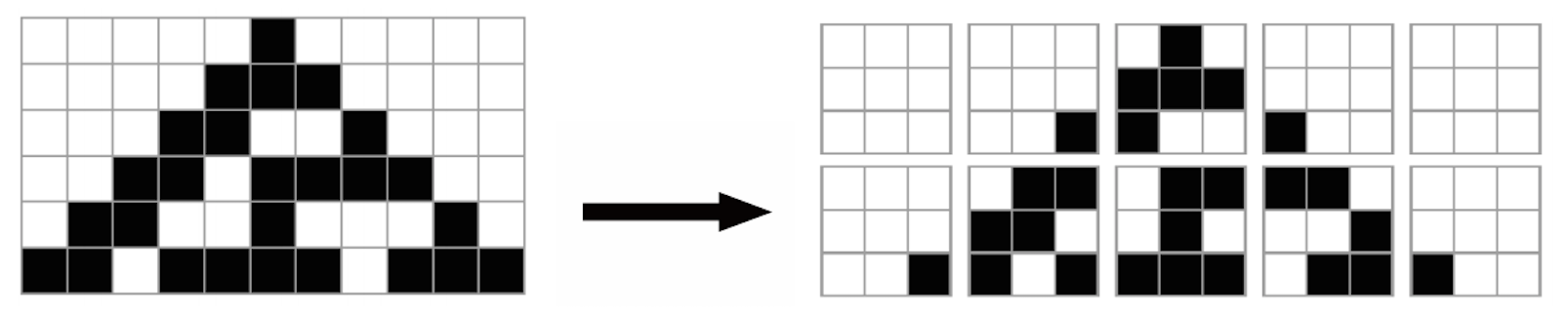}}

\caption{\label{bdmmapping}Algorithmic Information Dynamics (AID) guides the exploration and matching. Together CTM and BDM are the methods at the heart of AID~\cite{d4,d5,bdm}, helping to mine and guide the search for computable candidate models reproducing a piece of data (or the components of a smaller piece of data) from the space of all computer programs. Automacoin is a cryptocurrency, currently under development, that illustrates the way in which more responsible computer power use can help science using methods like CTM and BDM. (Bottom) The process of block decomposition of a ECA (or any other system) consists in dividing the CA time-space evolution into smaller blocks for which estimations of algorithmic randomness are precomputed using the Coding Theorem Method. This image is licensed under a Creative Commons Attribution 3.0 License.}
\end{figure}

The methods that I have developed over the years, building upon the theory of algorithmic complexity, can help map the space of computable models and provide the means to access this knowledge by mapping the space of  computable models and observations (data), making them natural underlying candidate mechanistic models of that data (observations). To this end, data is chunked in smaller pieces as illustrated in Fig.~\ref{bdmmapping}, with each piece having a corresponding candidate computable model from the space of computer programs.

Because the computer power needed to traverse the space is huge (as difficult as the most difficult computable function, the so-called Busy Beaver problem~\cite{rado}), we have used supercomputers in the past and have now proposed also new methods based on cryptocurrencies that can be exploited to compute the space, rather than computing useless one-way functions as it is typical for cryptocurrencies. 

The resulting algorithmic landscape in Fig.~\ref{bdmmapping} is not flat because different pieces of data have associated computable models (computer programs) of different lengths, so some parts of the data are more difficult to produce than others, i.e. require more information to be specified, either because they are disconnected from the rest or are parts in contact with other systems injecting new information.

Randomness may sometimes seem to be dominating science, because if we separate noise from signal it is what appears as noise that bothers science and is the subject of current investigation. Once science tames noise, it becomes signal, and science moves to the next apparent source of noise.

Science has provided us with all sorts of models, such as statistical and mechanistic models. Mechanistic models are very important because they suggest underlying mechanisms as candidates for what may occur with actual phenomena. A mechanistic/algorithmic approach provides a causal model from first principles. Unlike a statistical description, it is prescriptive, as it provides guidance for manipulating causes.

For example, in Fig.~\ref{sciencealgs}, a mechanistic model of the solar system is shown. In a mechanistic model such as this one, if we were forced to place the sun and the moon at a certain distance from each other in order to reproduce solar and lunar eclipses, then such a model would suggest a distance that can be verified or falsified by other means, that in turn may feed back to our model to make it better. The traditional approach is to consider a model that starts from certain values, and after some time provides some clue as to where the system may be at a future time, with some precision and in less time than the time taken by the actual unfolding of the phenomenon being studied. If a model does not comply with these requirements then it is traditionally seen as less valuable, so in general, the more precise and the faster it can be, the better. But more importantly, a model is better if it can explain more with less.

\begin{figure}[ht!]
\centering
\scalebox{.55}{\includegraphics{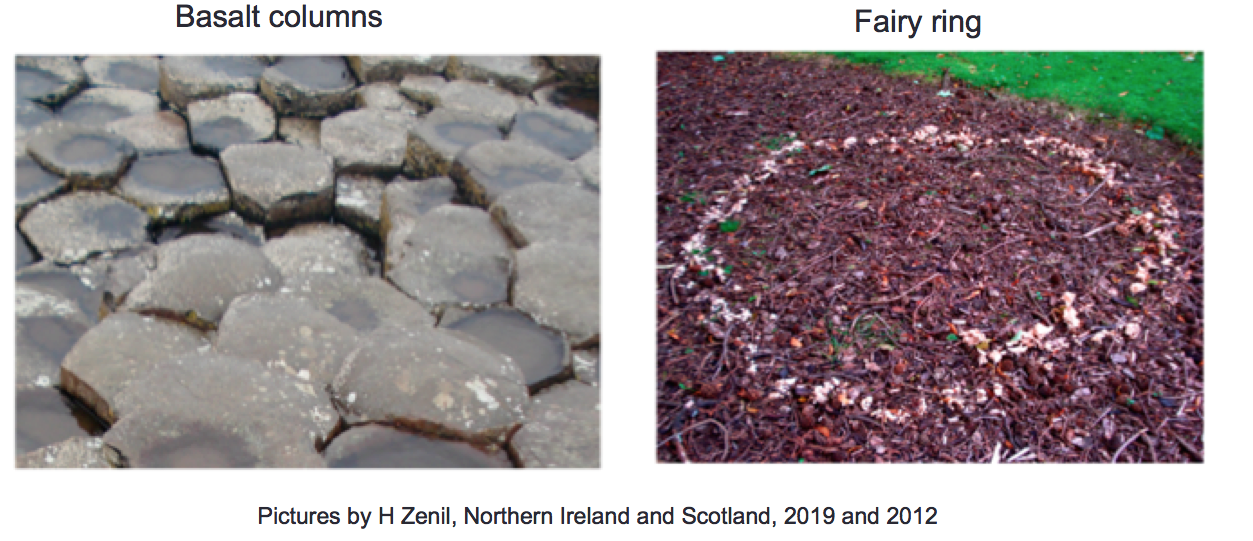}}
\caption{\label{naturepatterns}Patterns all over the surface of the planet made by living organisms other than humans and far removed from chance, such as fairy rings, basaltic hexagonal columns, termite mounds. Left: The Giant's Causeway in Northern Ireland. Hexagonal basalt columns formed by sea
water cooling incandescent lava relatively quickly, around 60 million years ago, as a result of the interplay of a small number of forces. Right: A `fairy ring' on the ground is formed in a natural manner by the differential growth of Micelios mushrooms, i.e. an extremely simple mechanical process. This image is licensed under a Creative Commons Attribution 3.0 License.}
\end{figure}

The movement of planets initially appeared to us as ungoverned, and some of them even seemed to be randomly moving in the sky, and were thus called \textit{wanderers} in Greek. It was not until Galileo, Copernicus, Kepler and later Newton that it was found that planets followed rules. But there were some movements of the internal planets such as Venus, as well as other celestial phenomena, which could not be fully explained by Newtonian mechanics, and thus General Relativity was needed---another mechanistic model to explain what again was apparently random but was found not to be so under the new model.

It is true that moving from one theory to another reveals that we cannot ever be sure that a given theory definitively explains the data, nor can we say that these rules or differential equations are followed by nature or physics, yet it is also undeniable that each iteration of a model is more encompassing, able to explain not only previously puzzling phenomena but many others as well. In this sense, the compass of what new theories explain, generally speaking, exceeds their own lengths. In other words, we gain much more explanatory and even predictive power with an increasingly compact number of theories over time. This is not only a clear indication that our world and universe is removed from randomness, with a handful of models being able to explain almost every aspect of physical experience, if not human experience, it is also strong evidence of the algorithmic nature of the world. Indeed, that the universe can be explained with theories and models of apparently ever-decreasing length means that the data they describe is formally of low algorithmic randomness, and that finding ever smaller models among all possible computable models is algorithmically very likely, which in turn, by the coding theorem, means that the world is algorithmically simple.

\subsection{Pervasiveness of Turing universality}

While it was known that Turing universality could be achieved with extreme simplicity (in terms of resources, i.e. state+symbols), it was not known how pervasive it was. Once a computer program is Turing universal it can simulate any other computer program. Our work  is the first to shed some light on quantifying how pervasive Turing universality may be. It turns out that the number of Turing-universal machines is of density one, meaning that basically almost all computer programs are capable of universal computation, given the appropriate initial conditions~\cite{juergenzenil2}. We have also shown how computer programs previously thought to be essentially different can actually simulate other computer programs of unbounded complexity. In this sense, we have shown a complete collapse of the so-called Wolfram classes in a fundamental way~\cite{juergenzenil2}. But we have also shown that there is an asymmetry in the number of possible compilers accessible during a given exhaustive compiler exploration time for performing these simulations, 
which reintroduces a hierarchy and re-establishes the classes from an epistemological perspective. The fundamental collapse, however, strengthens Wolfram's so-called Principle of Computational Equivalence. By way of the aforementioned asymmetry we also defined a topological measure based upon simulation networks, establishing the degree of difficulty involved in finding the right compilers to make a program simulate other programs. All this constitutes a completely novel Bayesian approach to Turing universality and a powerful framework within which to reprogram a system's behaviour. These ideas are also consistent with my many first views on programmability, based on Turing's 'imitation game,' for testing black box capabilities from input perturbation (interrogation) and output assessment. We propose to use the tools of algorithmic complexity to study such mappings as an ultimate behavioural evaluation framework.

We have conceived these tools based on the work of Chaitin and the other founders of AIT as alternatives to compression, tools which allow us to find underlying models. The methods represent a way to navigate through the space of all possible (small) computer programs in order to match computable models with data observations from the real world.

My tools based on Chaitin's work are helping us understand some aspects of dynamic systems useful in the context of living systems, and moreover able to steer and manipulate their behaviour.

\begin{figure}[ht!]
\centering
\scalebox{.6}{\includegraphics{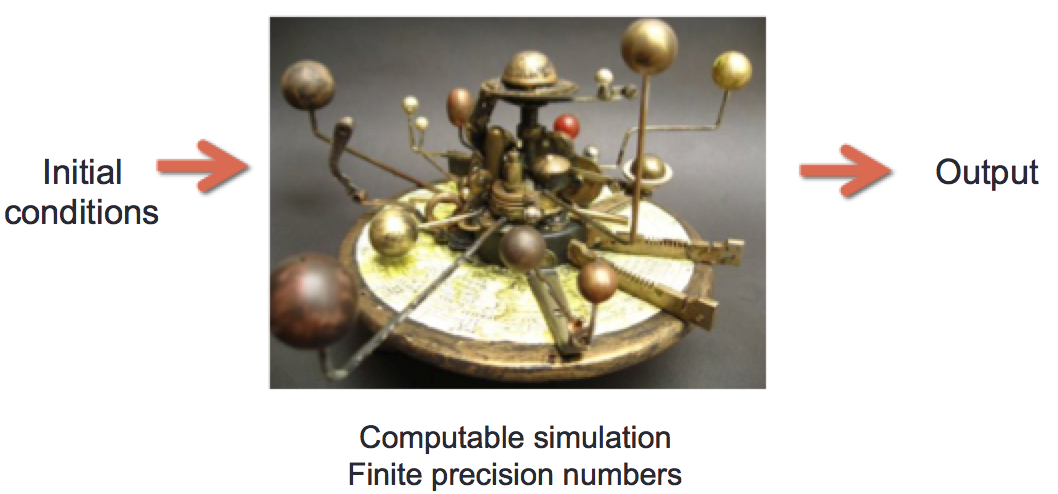}}
\caption{\label{sciencealgs}A mechanistic model is a model that can be run and  simulated by mechanical means by setting it in an initial condition, following a set of instructions, and determining an output, e.g. the state of the model. In this case, for example, the placement of the planets. Image from Wikipedia under Creative Commons license.}
\end{figure}

Another question of relevance concerns computability versus uncomputability. Here again, even assuming that there are noncomputable processes, our best tools consist in simulating everything in a digital computer by representing systems and their respective models with computable algorithms, even in those cases where we are dealing with apparently continuous differential equations, which, thanks to CPU's finite precision capabilities, in the end become discrete and symbolic approximations of these apparent and convenient continuous representations (see Fig.~\ref{sciencealgs}).

We have seen how we have contributed to matching data with computable models~\cite{maininfo}, but how successful and relevant are computable models? The distinction between models as equations and models as computer programs represents a gap between the discrete and the continuous that does not seem necessarily fundamental (see Fig.~\ref{sciencealgs}). While mechanistic models can also be wrong, they have the advantage of conforming to the laws of physics on the basis of which they are constructed, depending on the nature of the model, whether classical or not. While the model may look analogue because it moves in apparently continuous space and time, measurements are of finite accuracy. Most models in science today are not only mechanistic but algorithmic, and they run on electronic digital computers. They are thus computable models of the physical phenomena they are meant to describe. In other words, modern science has become a computational science, both in practice and in principle in many cases, and its power has not diminished. On the contrary, it has never been more powerful, in both its explanatory and its predictive capacity.

\section{Complexity in Nature}

\subsection{Algorithmic probability in biology}

I aim to understand the behaviour of artificial and biological systems from a computational perspective, and to develop methods to reprogram biological cells at a molecular/genetic level as we do computers. I do this by combining the power of the conceptual and methodological advances represented by the formal theories of information (Shannon, Kolmogorov, Chaitin, Solomonoff, Levin, et al) and computation (as advanced by Turing, Church, Kleene, et al), and by exploiting the versatility of principles from areas such as dynamical systems and complex networks, to develop a novel, elegant, and very sophisticated framework and field, namely algorithmic information dynamics, to help deal with questions of causation and produce model-driven, generative and mechanistic explanations for natural and artificial processes.

Biology is mostly discrete, without much room for apparent continuity. Not only is the genetic code discrete, finite, and composed of 4 letters, genes produce an integral number of viable proteins that either dock or do not, and genes can be simulated closely by modelling them as being either on or off. Proteins are very finite objects. They converge in great numbers, and when they don't they most likely never dock or fulfil their purpose. Chemical concentrations that appear continuous are only a useful representation. Living systems, however, are, in a manner of speaking, assembled out of chemical LEGO pieces. Even signals among cells are proteins that can be counted and followed one by one. Living systems themselves depend on the very binary choice that enables cells to differentiate between themselves and everything else by way of their perfectly limited membranes. Communication among cells is through the physical interchange of proteins---the same kind of LEGO pieces are involved here as everywhere else. And just as with everything else, in most regular cases the piece either docks and fulfils its purpose or it does not. Even Brownian motion that plays a role in, e.g., protein folding, is produced by finite and well-defined particles or molecules interacting with other molecules.

We have explored how non-trivial deterministic systems can display unbounded changes in complexity. We have proved that undecidability is essential for legitimate forms of OEE definitions and that this is also deeply connected to open and closed systems, with self-referential state-dependent systems displaying a greater degree of versatility. For OEE to be possible (under reasonable definitions), then open and undecidable dynamic systems are necessary. But even more important are the results that we have reported regarding the substitution of uniformly distributed mutations for mutations occurring according to the so-called Universal Distribution that predicts the way in which rule-based systems behave according to the computational power of their source (see paper J49). It turns out that this simple and sound substitution (because neither the world nor physical laws are truly random) can significantly speed up evolutionary convergence, may justify the need of genetic memory and modularity (e.g. gene organization), and even explain phenomena such as diversity explosions and mass extinctions that have no extrinsic (e.g. climate) explanations. 

Equipped with the measures that we have developed, my team and I aim to tackle a fundamental challenge in science: that of developing tools for causal discovery~\cite{nmi,maininfo}. This in order to unveil the design principles and generating mechanisms of arbitrary dynamic systems. In particular, the development of an interventional calculus based upon the theory of computability and algorithmic probability that identifies the key markers in genetic networks with which to steer and manipulate cell function and cell fate. The range of application of this work is very general, and aims to generate effective intervention tools to steer the causal content and thus the fate of artificial and natural complex systems.  I am engaged in devising 
strategies to understand, test and reprogram artificial and natural systems as models of computation. I have also devised two measures for testing the `algorithmicity’ and programmability of a system. We also introduced a perturbation-based calculus to study the information landscape of complex systems and networks---a calculus capable of unveiling information on the energy landscape---using a measure of algorithmic (re)programmability that connects the theory of dynamic systems with (algorithmic) information theory to reveal the dynamic landscape at the heart of the problem of causality. I use computer programs as models of the world. Determinism from classical mechanics implies that everything can be seen as a computer program and that the complexity of systems led purely by causal elements included in their description are dominated by their evolving time. This allows us to find clues to move systems between different functions,  thereby effectively reprogramming them~\cite{maininfo}.

\subsection{Algorithmic intelligence}

The human mind is algorithmically biased to suit specific purposes. Some of these algorithmic mechanisms can replace previously considered biases based on prior experience (e.g. the so-called system 1). We have known that random generation tasks, in which people have to generate random-looking sequences, are linked to some important aspects of mental and cognitive capacities. We have introduced a new approach to the study of the behaviour of animals and humans based on algorithmic information theory. We have shown that humans best outsmart computers when they are tested on randomness generation at the age of 25~\cite{ploscompbio}. When competing against all possible algorithms, we have found that human behaviour is at its most algorithmically random at 25 years of age, thereby introducing a new measure of cognitive complexity possibly related to other biological and cultural phenomena, such as creativity. Indeed, humans produce more randomness than most computer programs when they are 25 years old. In parallel, we are also working on measures of integrated information with which to profile brain networks.

\begin{figure}[ht!]
\centering
\scalebox{.25}{\includegraphics{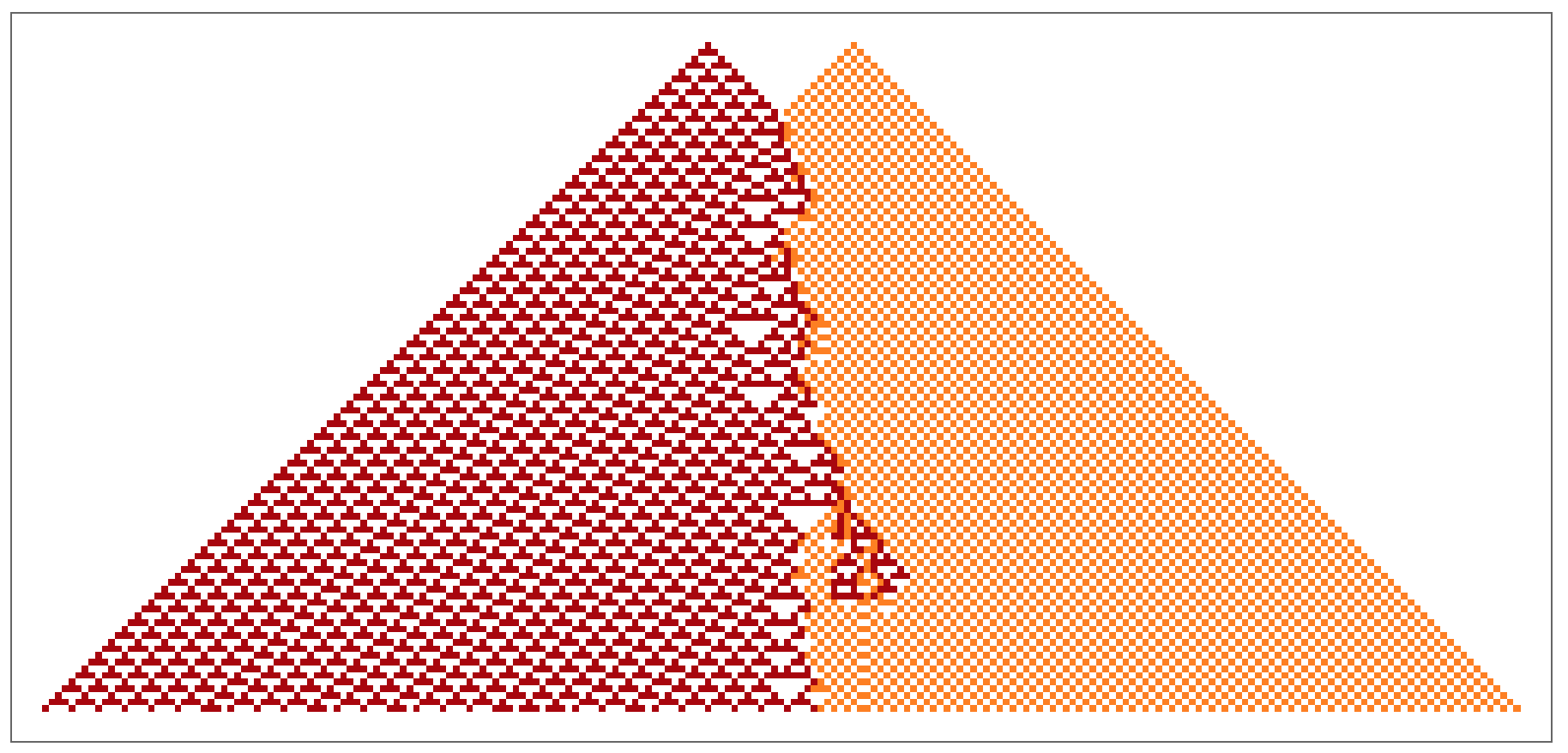}}

\bigskip

\scalebox{.25}{\includegraphics{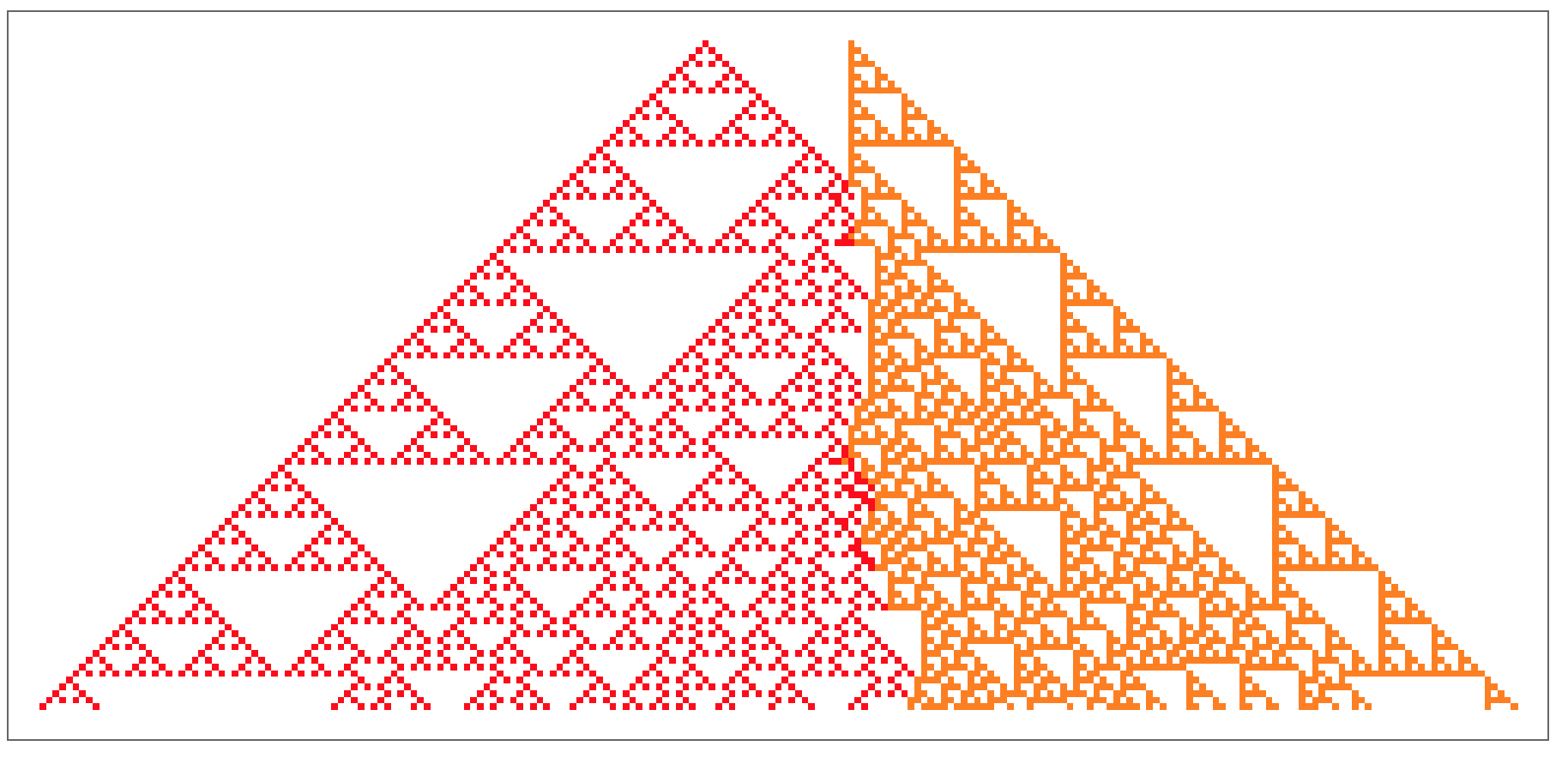}}

\caption{\label{rule54v50}Examples of simple interacting computer programs (top: ECA rules 54 and 50, bottom: ECA rules 82 and 60) with a simple function determining the interaction from which data emerges and gets convoluted. Different rules come from different generating sources. In this case different Elementary Cellular Automata rules. Observers will often face this problem and will not be able to disentangle data without the right tools. The world is made of this type of interaction, producing apparent complexity and concealing the underlying generating rules that new methods can help disentangle~\cite{nmi}. They do not even have to be complicated in themselves, but when interacting they produce a highly complex pattern that cannot be quantified with traditional tools, and may be irreducible.}
\end{figure}

In Fig.~\ref{rule54v50}, ECA can be seen to be interacting with each other and generating structures according to a meta CA rule. Often these interactions appear random, and to an observer they may be difficult to deconvolve or disentangle, but new methods introduced by my team and based on computability theory and the work of Chaitin and Levin (together with the work of Kolmogorov and Solomonoff) help new machine learning algorithms to distinguish the sources by their most likely causal generating mechanisms (in this case, each side is generated by either rules 54 and 50 or 82 and 60)~\cite{nmi}.

Most systems in nature are like black boxes that we cannot open and fully understand, and we are therefore left with approaches not very different from the black box approach of Alan Turing---his imitation game---for dealing with the challenge and question of machine intelligence. What we conceived was a test for systems where questions would be inputs and answers would be compressed. One could then discern the variation among the compressed outcomes for answers of varying complexity and assess the system's capabilities. The idea has evolved into a perturbation algorithmic calculus able to characterize artificial and biological systems~\cite{maininfo,nmi}.

These tools are becoming a new approach to the whole area of machine learning and even of Artificial Intelligence, introducing the inferential power of algorithmic probability to complement the current state of areas such as deep learning, which are weak on the fundamentals for understanding human intelligence, such as symbol manipulation, logic, inference, and the understanding of cause and effect.

\subsection{An algorithmic universe}

\begin{figure}[ht!]
\centering
\scalebox{.45}{\includegraphics{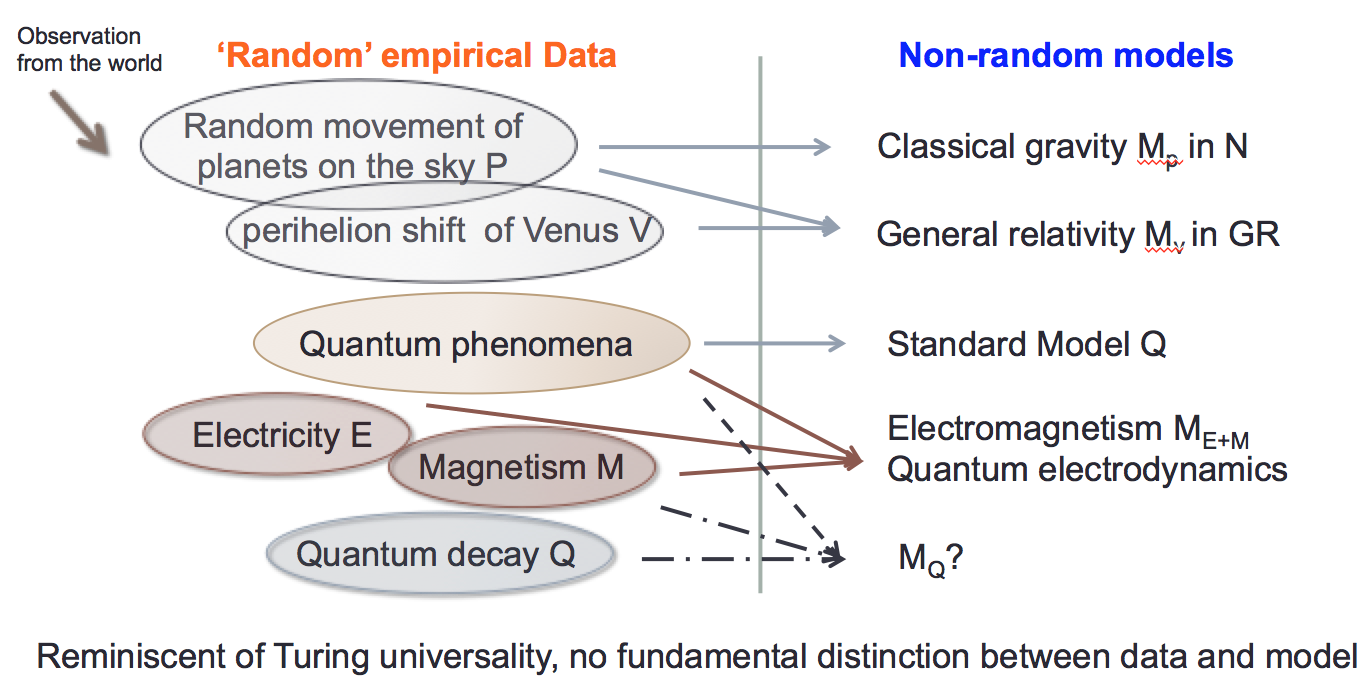}}
\caption{\label{movingtarget}Science moves apparently random data from observation to model/theory. Indeed, scientific models tend to cover previously random data together with other previously unexplained phenomena. Whether there is a model that can encompass all other models is an open question, but the tendency so far has been clear, and not an artefact of humans as intelligent or conscious observers. This image is licensed under a Creative Commons Attribution 3.0 License.}
\end{figure}

Certain parts of the universe seem ordered and structured (see Fig.~\ref{naturepatterns}). On Earth, for example, life is an example of organization and structure, contrasting with what can be deemed background noise (comparable to what appears on the screen of an old non-tuned analogue television) left over by the Big Bang and serving as proof of the state in which the universe found itself after its first moments of existence. What Turing would probably never have guessed is that when running random Turing machines, the machines produce highly structured objects. Could this be merely an interesting analogy or could it perhaps be an actual indication that the universe is more algorithmic than initially expected? If so, then Turing machines and computer programs are not just products of the human imagination, they are perhaps responsible for the order in the universe. Alan Turing may have had an intuitive answer to this question, as he was also interested in structure formation and helped found another area in biochemistry called \textit{morphogenesis}~\cite{turingmorpho}, the study of pattern formation, starting from a simple shape which would first break its symmetries at random.

We have made great progress at taming apparent randomness since ancient times, with mathematical logic and rationality to begin with---even if we sometimes seem to be regressing to ancient times. A long time ago we left behind explanations based on divinity, magic or paranormal phenomena in exchange for tools such as inference, and statistics. Approaches such as correlation in regression were very useful but they have been overused and abused, and new and better tools for dealing with modern causality are badly needed.

\begin{figure}[ht!]
\centering
\scalebox{.3}{\includegraphics{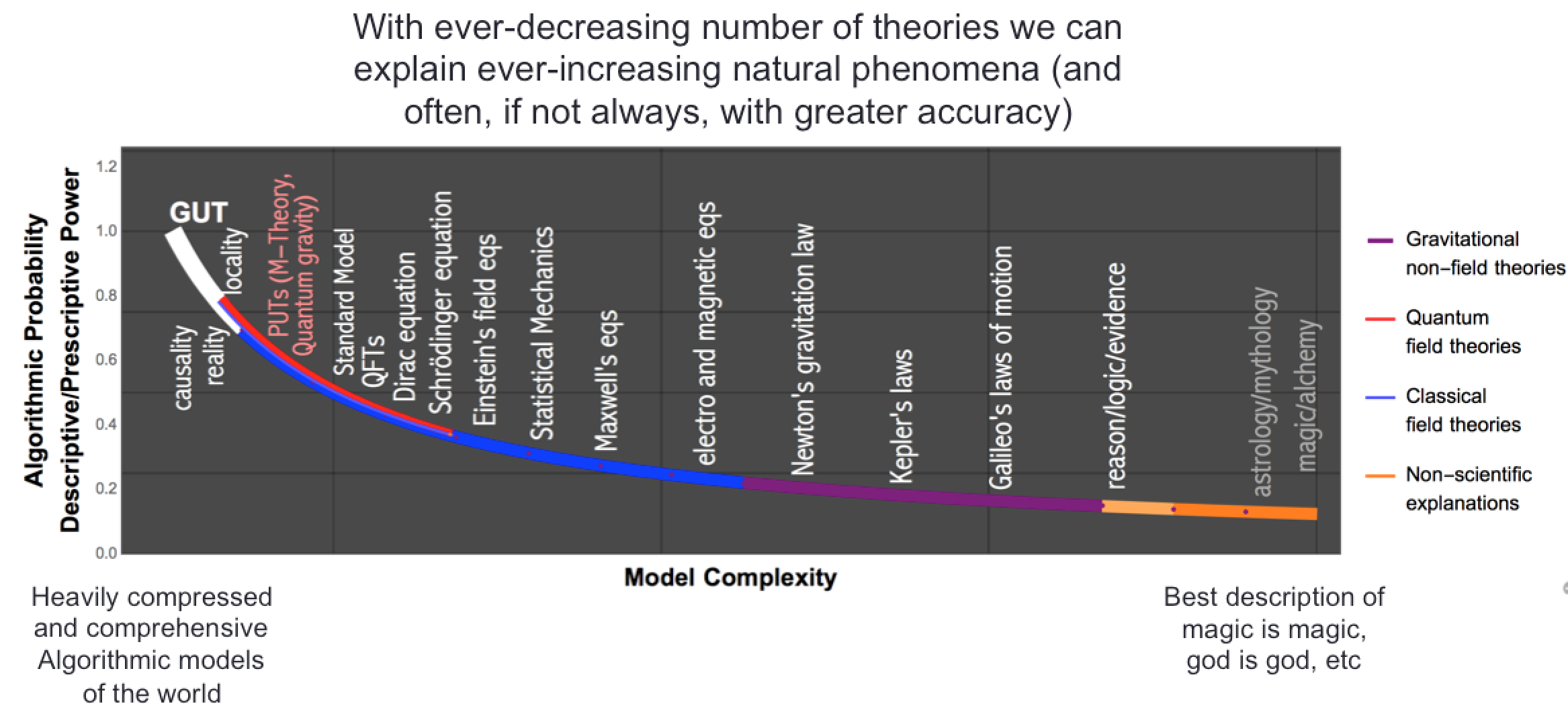}}
\caption{\label{gut}This diagram shows in what direction science has successfully been operating in previous centuries, and how this has been increasingly sped up in the last century, relying on a handful of fundamental theories that have unified seemingly disparate areas of science, each time explaining more with less. The existence of science is evidence of the algorithmic nature of the world, and the fact that more modern scientific models encompass larger observations strengthens the hypothesis of an algorithmic universe. This image is licensed under a Creative Commons Attribution 3.0 License}
\end{figure}

With ever-increasing predictive power, science moves random observations under the explanation of computable models and merges previously computable models into ever more encompassing models whose program-size may be difficult to quantify but whose explanatory power is increasingly greater, hence indicating a clear pattern whereby the universe looks increasingly ordered and of low Kolmogorov-Chaitin complexity (Fig.~\ref{gut}). Science has been mostly motivated by models, has been a model-driven practice. A case can be made regarding the size of new models/theories. For example, taken at face value, the description of the equations for General Relativity (GR) may give the impression of being longer than those for Newtonian classical gravity, hence a possible indication (upper bounds of comparable measure) of their underlying algorithmic complexity. However, the GR equations are strictly shorter than the equations specifying the Newtonian version of gravity plus all the corrections that they require to account for the phenomena that they cannot account for by themselves, such as the movement of the internal planets, in particular the rate of precession of the perihelion of Mercury's orbit. In comparison, General Relativity requires few to no corrections. In other words, Newtonian mechanics would require a stream of regular adjustments to match the computable accounts of General Relativity for such phenomena.

One may also argue that even GR can only predict the precession movement of Mercury for a certain time because of 3-body kinds of phenomena, and this is right, but the fact that we can predict with extremely high accuracy the movements of Mercury for the next 1000 years is different from not been able to account for its movements at all using Newtonian mechanics.

So, in some sense, science can be illustrated as the practice of moving natural phenomena from what we as observers perceive as random, towards non-random phenomena. Science moves empirical data from apparent randomness to algorithmic non-randomness (Fig.~\ref{movingtarget}). Each one of these phenomena under the non-random quadrant was previously on the random phenomena quadrant. One can see that there is also a cascade effect in the right quadrant, with models explaining all or several other previously advanced models. So the overwhelming evidence is that the world has a strong algorithmic component, because science can explain, if not predict, most of what happens in the universe to an increasingly high degree of accuracy and comprehensiveness. The fact that these models compress more observations from natural phenomena in a reduced number of models is an indication that, indeed, compression is comprehension.

Even in some non-classical interpretations noise may be only apparent, because in fact all deterministic trajectories are explored and we just happen to be in one particular branch of a multiverse. That of course does not solve the problem of the source of global randomness at the level of the multiverse itself, determining why we experience only one random universe and not all others. However, evidence (Fig.~\ref{movingtarget} and Fig.~\ref{gut}) also suggests that what we think is noise is often a signal whose source is unknown or irrelevant to a system of interest. 

Time has tell again and again that every time there is two seemingly disparate phenomena they are often sides of a common underlying duality or symmetry, and that every time that there is a physical constant associated to an apparent fine-tuned irrational number appearing as a hyper-parameter of our universe, there is an algorithmic model in which such a constant is emergent from first principles as it has happened in areas such as optics or electromagnetism. Their presence suggest simpler encompassing underlying computable models able to collapse seemingly fundamental constants and connect apparent disparate phenomena.

\section{Conclusions}

Studying randommness from the computing perspective affords us a framework for studying the nature of the world that discerns the way in which patterns in the universe are distributed. In a world of computable processes in which the laws (like programs) do not have a slantwise distribution, $AP(s)$ would indicate the probability of a natural phenomenon occurring as a result of running a program. Distribution $AP(s)$ has an interesting particularity: it can start from basically anything, and, like a robust distribution, it remains qualitatively unchanged. It is the process that determines the form of the distribution and not the initial distribution of the programs. This is important because one does not make any strong initial assumption about the distribution of the initial conditions or of the laws of physics.

Computer programs can be looked at from a certain vantage point as laws of physics. If one starts with a random initial condition (input) and executes a program chosen at random, there is a very good probability that its final appearance will be regular, and frequently very well organized. By the same token, if one were to shoot out particles at random, the probability of groups forming in the way they do would be so small---in the absence of laws of physics---that nothing whatsoever would happen in the universe. Perhaps Alan Turing hasn't only helped us understand the world of computing machines and computing programs, founding an entire scientific area, but has also taught us a great deal about the universe in which we live and how it came to be the way it is. Algorithmic complexity and algorithmic probability may explain the unreasonable effectiveness of mathematics in the natural sciences, showing it to have been based on very solid foundations.

\end{document}